# Attainability of high velocities for impact ignition

F.Winterberg

University of Nevada, Reno

USA

October 2009

**Abstract**

To reach the flyer plate velocities in excess of 1000km/sec required for impact ignition, it is proposed to combine the ablation acceleration of a dense hydrogen jet by its isentropic compression in a convergent Prandtl-Meyer flow, magnetically insulated by the Nernst effect against the wall confining the flow to reduce friction losses. A flyer plate placed at the front of the flow can there be accelerated to much higher velocities.

# 1. Introduction

The idea of impact ignition in uncompressed liquid deuterium-tritium (DT) goes back to a paper by the author in 1963 [1], where it was proposed to reach the ignition temperature of DT by a beam of small solid particles, electrostatically accelerated to 1000km/sec. To reach upon impact the required temperatures of ~ $10^8$ °K, the size of the particles had to be larger than the mean free path at $10^8$ °K, a condition for the creation of a shock wave upon impact.

The principle difficulty of this idea was that for launching a thermonuclear detonation wave in liquid DT, the particles would have to be of cm-size dimensions, or a beam of such particles would upon convergence act like one particle of this size.

Most recently the idea was revived in a paper by Murakami and Nagatomo [2], who suggested the fast impact ignition could be achieved by the laser ablation acceleration of a small flyer plate onto a highly compressed DT target, where a single much smaller particle could cause ignition. With this idea, experimentally observed fusion neutron yields could be increased 100-fold [3]. The velocities reached in this experiment where about 600km/sec, still short of the over 1000km/sec needed for impact ignition. It is the purpose of this note the explain how higher velocities, hopefully well over 1000km/sec can be reached with a modified concept.

## 2. The Murakami-Nagatomo concept

The concept proposed by Murakami and Nagatomo is shown in Fig.1. It adopted a configuration proposed by Kodama [4], where a cone is stuck into a DT pellet, facilitating the access of the laser fast ignition pulse to the center of the DT pellet. For the fast impact ignition of the highly compressed DT pellet, a flyer plate placed in the cone was ablatively accelerated by a nanosecond laser pulse.

The important advantage of any impact fusion concept is that if permits to cumulate kinetic energy into projectile more slowly, because the acceleration can take place over a large length. A slower acceleration may take more energy, but it is the high final velocity at a sufficiently large energy, which is important for impact ignition.

## 3. Modification of the Murakami-Nagatomo concept to reach higher velocities

In the Murakami-Nagatomo concept a much higher neutron yield was reached, even below the critical velocity in excess of 1000km/sec needed for ignition. A higher velocity would require larger laser energy, but a higher velocity, leads to increased friction and radiation losses of the flyer plate touching the wall of the cone, which in the proposed modified version can hopefully be substantially reduced.

The proposed modified configuration is shown in Fig. 2. In it the cone is replaced by a convergent Prandtl-Meyer compression duct, where for a supersonic flow all the Mach lines converge into one point, isentropically compressing the flow [5]. The flow is ablatively launched into liquid hydrogen a laser or particle beam. But not only have the Mach lines converge in space onto one point, they have also converge there in time by a programmed laser or particle beam [6]. At the beginning the hydrogen will have a low temperature, but soon rises above $10^5$ °K, where it is transformed into fully ionized plasma.

With an externally applied axial magnetic field, serving as a seed field to generate in the boundary layer between the Prandtl-Meyer flow and its confining wall a toroidal current by the thermomagnetic Nernst effect, amplifying the externally applied magnetic field, the flow is repelled from the wall. For a hydrogen plasma in the boundary layer near the wall $n^4T$=const, instead of nT=const. as it would be in the absence of Nernst effect [7]. Without the Nernst effect where nT=const, the plasma density near the wall, where T→0, becomes large and with it the friction of the flow against the wall. For $n^4T$=const. the plasma density near the wall is reduced. The Nernst effect acts here in a similar way as the insulating hot steam layer between the surface of a hot plate and a drop of water placed on the hot plate, in what is there known as the "Leidenfrost effect".

The Nernst effect also drastically reduces the bremmtrahlungs-losses near the wall. These losses are given by

$$\varepsilon_r = \text{const.} \, n^2 \sqrt{T} \qquad (1)$$

For nT=const. they go in proportion to $T^{-3/2}$ as T→0 in approaching the wall, while for $n^4T$ they remain constant.

The Nernst effect thus acts like an efficient lubricant of the flow against the wall, and the same applies to the flyer plate which by this lubricating effect can be expected to reach much higher velocities than would otherwise be possible.

Let us assume that the ablation temperature the liquid hydrogen is of the order ~$10^6$ °K, implying an ablation velocity $v_A$ ~$10^7$ cm/s. Inserted into the equation rocket equation

$$v = v_A \ln(m_0/m_1) \qquad (2)$$

and assuming a mass ratio $m_0/m_1$~30, which is one third of the volume of the Prandtl-Meyer duct, this leads to a velocity v ~ 700km/sec. In the final segment of the convergent duct the velocity increases in the inverse proportion to the radial distance from the center of convergence, at least by a factor 2, sufficient to reach the more than required 1000km/s.

**Figures**

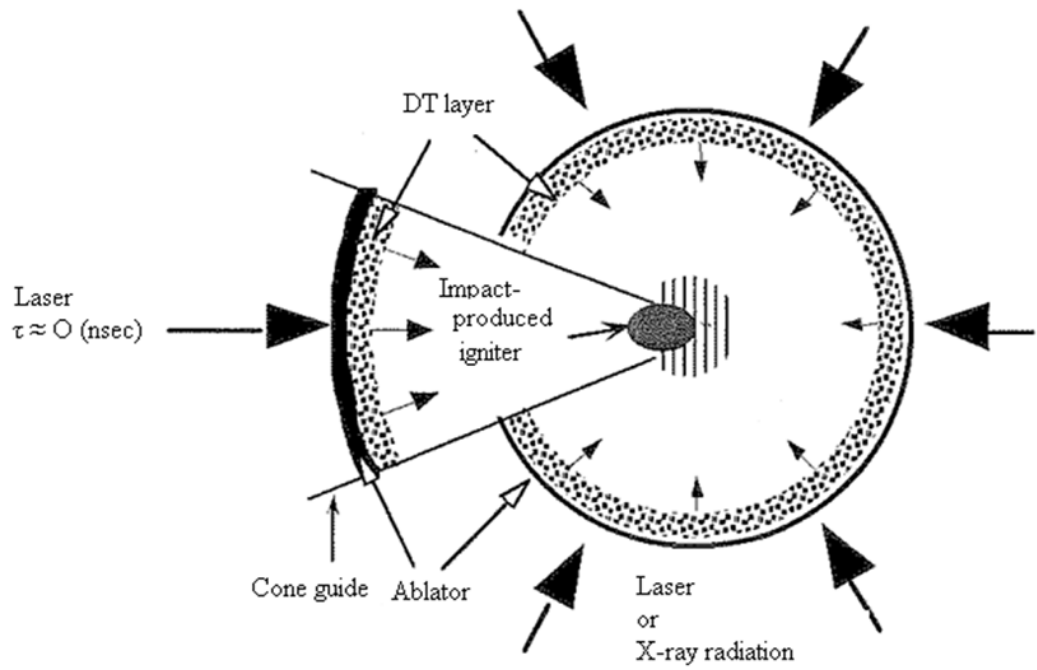

Fig.1 The original Murakami-Nagatomo impact ignition concept.

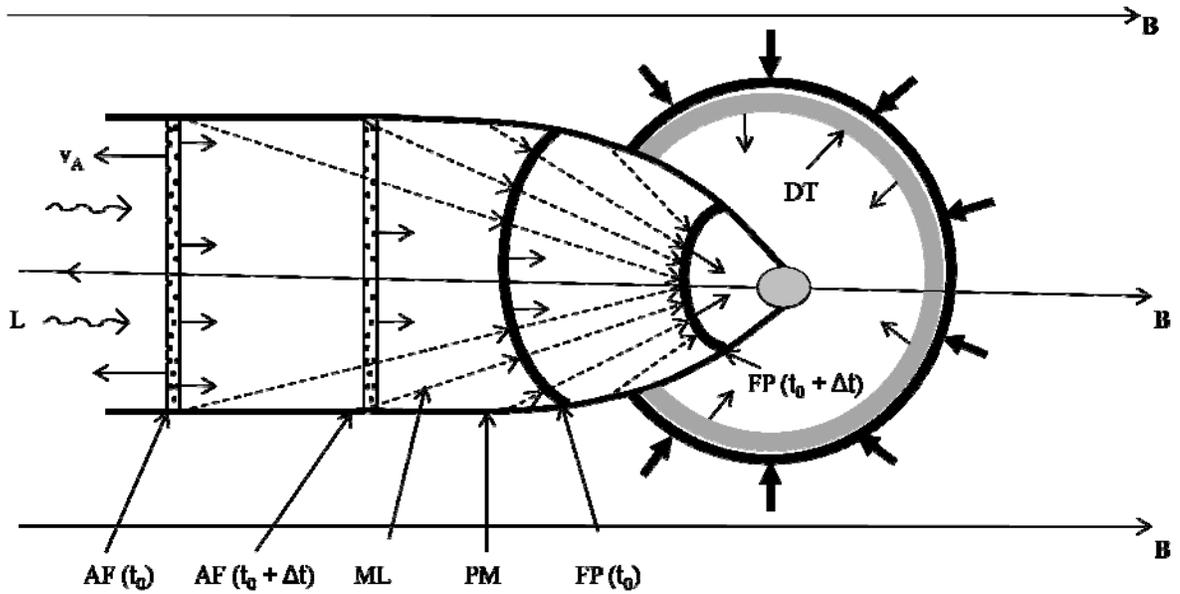

**Fig.2** The modified impact ignition concept L laser or particle beam; $v_A$ ablation velocity; AF ablation front at $t=t_0$, and $t=t_0+\Delta t$; ML Mach lines; PM Prandtl-Meyer isentropic compression flow duct; FP flyer plate at $t=t_0$, and $t=t_0+\Delta t$; **B** axial magnetic field; DT deuterium-tritium.